\documentclass[12pt]{amsart}
\usepackage{amssymb}
\usepackage{amsmath,amsthm,amsfonts}  
\usepackage[applemac]{inputenc} 
 \usepackage{graphicx}
\usepackage{amsthm}
\usepackage{bbm}
\usepackage{ulem}
\usepackage{verbatim}
\usepackage{pdfsync}
\usepackage{fancyhdr}
\usepackage[default]{frcursive}
\usepackage[T1]{fontenc}
\usepackage{float}
\usepackage{color}
\usepackage{xcolor}
\usepackage[breaklinks,colorlinks,backref]{hyperref}
\hypersetup{
    colorlinks, 
    linktoc=all, 
    linkcolor=red,  
  citecolor=hyptxt,
  urlcolor=blue}
  \hypersetup{
  citebordercolor=,
  filebordercolor=green,
  linkbordercolor=blue
}
\definecolor{hyptxt}{rgb}{0.7, 0.4, 0.9}
\usepackage{bibtopic}
\DeclareFontFamily{U}{mathx}{}
\DeclareFontShape{U}{mathx}{m}{n}{<-> mathx10}{}
\DeclareSymbolFont{mathx}{U}{mathx}{m}{n}
\DeclareMathAccent{\widehat}{0}{mathx}{"70}
\DeclareMathAccent{\widecheck}{0}{mathx}{"71}

\usepackage[full]{textcomp} 
     \usepackage{fbb} 
     \usepackage[scaled=.95]{cabin}
     \usepackage[varqu,varl]{inconsolata}
     \usepackage[libertine,bigdelims,vvarbb]{newtxmath}
     \usepackage[cal=boondoxo]{mathalfa}
     \usepackage[T1]{fontenc}
\useosf

\newcommand{\red}[1]{\textcolor{red}{#1}}
\newcommand{\ii}{\mathsf{i}}
\def\calH{{\mathcal H}}

\def\R{\mathbb{R}}
\def\N{\mathbb{N}}
\def\C{\mathbb{C}}
\def\Z {\mathbb{Z}}

\def\lg{\langle }
\def\rg{\rangle }
\def\deq{\stackrel{\mathrm{def}}{=}}

\def\vap{\mathfrak{w}}

\def\ud{\mathrm{d}}

\def\sfM{\mathsf{M}}

\def\bu{\mathbbm{1}}

\numberwithin{equation}{section}

\begin{document}


%
%

\title[Weyl-Heisenberg universality]{UNIVERSALITY OF THE WEYL-HEISENBERG SYMMETRY AND ITS COVARIANT QUANTISATIONS}

\author[Gazeau]{JEAN-PIERRE GAZEAU}

\address{ Astroparticule et Cosmologie, CNRS, Universit\'e Paris Cit\'e, \\ Paris, F-75013, France}
\email{gazeau@apc.in2p3.fr}

\author[Habonimana]{CELESTIN HABONIMANA}

\address{Ecole Normale Sup\'erieure du Burundi,\\ Bujumbura, Burundi}\email{habcelestin@yahoo.fr}

\author[Murenzi]{ROMAIN MURENZI}

\address{Dep. Physics, Worcester Polytechnic Institute,\\ Worcester, MA 01609, USA}
\email{rmurenzi@wpi.edu} 

\author[Zlotak]{AIDAN ZLOTAK}

\address{Dep. Physics, Worcester Polytechnic Institute,\\ Worcester, MA 01609,USA}
\email{ahzlotak@wpi.edu}

\maketitle


\begin{abstract}{The Weyl-Heisenberg symmetries  originate from translation invariances  of various manifolds viewed as phase spaces, e.g. Euclidean plane, semi-discrete cylinder, torus, in the two-dimensional case, and higher-dimensional generalisations. In this review we  describe, on an elementary level, how this symmetry emerges through displacement operators and standard Fourier analysis, and how their unitary representations are used both in Signal Analysis (time-frequency techniques, Gabor transform) and in quantum formalism (covariant integral quantizations and semi-classical portraits). An example of application of the formalism to the Majorana  stellar constellation in the plane  is presented.} 
\end{abstract}


\tableofcontents

\section{Introduction}
\label{intro}

The freedom to choose the origin in the Euclidean plane allows for significant and complex advancements, a concept that serves as the leitmotif of this work, which is dedicated to the Weyl-Heisenberg symmetry \cite{heisenberg25,weyl28,vonneumann32,perel86} and its numerous applications, like time-frequency signal analysis and quantum mechanics or optics as it is shown in \cite{gabor46,feichtinger98,flandrin98,grochenig01,feichtinger03,flandrin18,feichtinger09,klimov09}, and more specifically in  our recent works \cite{CIQ12,CIQ8,CIQ20,CIQ21,fabre23,CIQ22,CS96,CIQ25,CIQ18}. This symmetry extends, in a non-commutative manner, the translation symmetry of the plane.

The structure of the paper is organized as follows: Section \ref{WHGrs} provides an elementary overview of the various Weyl, or Weyl-Heisenberg (WH), groups. In Section \ref{signal}, we discuss the pivotal role played by the unitary irreducible representations (UIRs) of the WH group in continuous time-frequency analysis of one-dimensional signals.

Section \ref{sdcyl} explores another form of WH symmetry, specifically for the so-called semi-discrete cylinder $\mathbb{Z} \times \mathbb{S}^1$, along with its UIRs. Section \ref{quantiz} introduces the general (de-)quantization framework and the procedures that will be applied in subsequent sections. This section covers integral quantization in Subsection \ref{iquantiz}, semi-classical portraits in Subsection \ref{sdport}, covariant integral quantization in Subsection \ref{CIQ} involving a group or one of its left cosets, and an example \red{of} construction in Subsection \ref{weight} that incorporates probabilistic elements using density operators.

As indicated, this general procedure is applied in Section \ref{WHintquant} to the case of the Weyl-Heisenberg symmetry. An application of our (de-)quantization formalism to the  Majorana stellar constellation in the plane is outlined in Section \ref{stellar}. The paper concludes by suggesting future directions for this work in Section \ref{conclu}.

\section{Weyl-Heisenberg groups}
\label{WHGrs}
\subsection{Lowest dimensional WH}
The additive group $\R$ is also the group $\mathrm{T}_\R=\{\R\ni a\,:\, x\in \R \mapsto x+a \in \R\}$ of translations on the line. It has the $2\times2$ matrix commutative representation
\begin{equation}
\label{MatTrR1}
a \leftrightarrow \begin{pmatrix}
   1   &  a  \\
   0   &  1
\end{pmatrix}\, , \quad \begin{pmatrix}
   1   &  a  \\
   0   &  1
\end{pmatrix}\begin{pmatrix}
   x  \\
     1
\end{pmatrix}= \begin{pmatrix}
   x + a  \\
   1
\end{pmatrix}\,. 
\end{equation}
Such a $2\times2$ matrix representation map can be extended to the additive group $\R^2\sim \mathrm{T}_{\R^2} $ if the latter is viewed as the additive group $\C$ of complex numbers, 
\begin{equation}
\label{MatTrC1}
C\in a \leftrightarrow \begin{pmatrix}
   1   &  a  \\
   0   &  1
\end{pmatrix}\in \mathrm{M}(2,\C)
\end{equation}
 In this case $0$ and $1$ are viewed as complex numbers. If we wish to get a real, commutative,  matrix representation of $\mathrm{T}_{\R^2}$ we have to raise, at least by 1,  the dimension of the representative matrix, for instance, 
 \begin{equation}
\label{MatTrR2}
\R^2\in (a,b)  \leftrightarrow \begin{pmatrix}
   1   &  0 & a \\
   0   &  1 & b\\
   0   & 0 & 1
\end{pmatrix}\, , \quad \begin{pmatrix}
   1   &  0 & a \\
   0   &  1 & b\\
   0   & 0 & 1
\end{pmatrix}\begin{pmatrix}
   x  \\
     y\\
     1
\end{pmatrix}= \begin{pmatrix}
   x + a  \\
   y + b\\
   1
\end{pmatrix}\,. 
\end{equation}  
Now, the alternative choice 
 \begin{equation}
\label{MatTrR2}
\R^2\in (a,b)  \leftrightarrow M_{(a,b)}=\begin{pmatrix}
   1   &  a & 0 \\
   0   &  1 & b\\
   0   & 0 & 1
\end{pmatrix}
\end{equation}
does represent  $\mathrm{T}_{\R^2}$ up to an extra matrix:
   \begin{equation}
\label{MatTrR2}
\begin{split}
M_{(a,b)}M_{(a^{\prime},b^{\prime})}&= \begin{pmatrix}
   1   &  a & 0 \\
   0   &  1 & b\\
   0   & 0 & 1
\end{pmatrix}\begin{pmatrix}
   1   &  a^{\prime} & 0 \\
   0   &  1 & b^{\prime}\\
   0   & 0 & 1
\end{pmatrix} = \begin{pmatrix}
   1   &  a + a^{\prime} & ab^{\prime} \\
   0   &  1 & b + b^{\prime}\\
   0   & 0 & 1
\end{pmatrix}\\ &= M_{(a+a^{\prime},b+b^{\prime})} + ab^{\prime} E_{13} \, .
\end{split} 
\end{equation}
With the standard notation for the elements of the canonical basis 
\begin{equation}
\label{Eij}
\left\{E_{ij}\, , \, \left(E_{ij}\right)_{kl}=\delta_{ik}\delta_{jl}\, , \, 1\leq i,j,k,l \leq n\right\}
\end{equation}
of the matrix algebra M$(n,\R)$, the term  $E_{13}$ is equal to $E_{13}=  \begin{pmatrix}
   0   &  0 & 1 \\
   0   &  0 & 0\\
   0   & 0 & 0
\end{pmatrix}$. This extra term makes the map \eqref{MatTrR2} noncommutative:
\begin{equation}
\label{ncWH}
\left[M_{(a,b)},M_{(a^{\prime},b^{\prime})}\right]=  \left(ab^{\prime}-ba^{\prime}\right) E_{13} = \mathbf{v}\wedge\mathbf{v}^{\prime} E_{13} \,,
\end{equation}
with  $\mathbf{v}=(a,b)$. We thus are led to deal with the group  of unit upper matrices of order $3$:
\begin{equation}
\label{WH1}
\left\{M_{(a,b)} + c E_{13} = \begin{pmatrix}
   1   &  a & c \\
   0   &  1 & b\\
   0   & 0 & 1
\end{pmatrix} \,,\, a,b,c\in \R\right\}\, . 
\end{equation}
They form a version of the so-called \textit{polarised} Heisenberg or Weyl-Heisenberg group PH$_1$: 
\begin{equation}
\label{WHintL1}
\left(M_{(a,b)} + c E_{13} \right) \left(M_{(a^{\prime},b^{\prime})} + c^{\prime} E_{13} \right) = M_{(a+a^{\prime},b +b^{\prime})} + (c + c^{\prime}  + ab^{\prime}) E_{13} 
\, . 
\end{equation}
It is a real nilpotent group.   Higher-dimensional group versions PH$_n$ of PH$_1$ are  immediate: 
\begin{equation}
\label{PHn}
\mathrm{PH}_n = \left\{M_{(\mathbf{a},\mathbf{b})} + c E_{13} = \begin{pmatrix}
   1   & {}^{\mathrm{t}}\mathbf{ a} & c \\
   0   &  \bu_n & \mathbf{b}\\
   0   & 0 & 1
\end{pmatrix} \,,\, \mathbf{a},\mathbf{b} \in \R^n,\, c\in \R\right\}\,, 
\end{equation}
with
\begin{equation}
\label{WHintLn}
\left(M_{(\mathbf{ a},\mathbf{ b})} + c E_{1n} \right) \left(M_{(\mathbf{ a}^{\prime},\mathbf{ b}^{\prime})} + c^{\prime} E_{1n} \right) = M_{(\mathbf{ a}+\mathbf{ a}^{\prime},\mathbf{ b} +\mathbf{ b}^{\prime})} + (c + c^{\prime}  + \mathbf{ a}\cdot\mathbf{ b}^{\prime}) E_{1n} 
\, . 
\end{equation}
We note that $E_{1n} $ commutes with any element of PH$_n$: the subgroup $C=\lambda E_{1n} \, , \, \lambda\in \R$ is the center of PH$_n$. 

Going back to PH$_1$, since $M_{(a,b)} + c E_{13}  \equiv M_{(a,b)} + c^{\prime} E_{13} $ mod $C$, it is natural to display the symplectic  commutator $\mathbf{v}\wedge\mathbf{v}^{\prime} E_{13} $ in the internal law  \eqref{WHintL1} by redefining a group  element as
\begin{equation}
\label{defWH}
(c,a,b):=  M_{(a,b)} + \left(c+ \frac{1}{2}ab\right)  E_{13} = \begin{pmatrix}
   1   &  a & c + \frac{1}{2}ab\\
   0   &  1 & b\\
   0   & 0 & 1
\end{pmatrix}
\end{equation}
in such a way that the internal  law now reads as 
\begin{equation}
\label{WHintL1}
(c,a,b)\,(c^{\prime},a^{\prime},b^{\prime})= \left(c+c^{\prime} + \frac{1}{2}\mathbf{v}\wedge\mathbf{v}^{\prime},a+a^{\prime},b+b^{\prime}\right)\,. 
\end{equation}
The unity is $e= (0,0,0)$ and the inverse is given by
\begin{equation}
\label{invWH1}
(c,a,b)^{-1}= (-c,-a,-b)\,. 
\end{equation}
The Weyl-Heisenberg (or  Heisenberg) group equipped with the law \eqref{WHintL1} is  denoted H$_1(\R)$, or simply H$_1$,  in the sequel.

\subsection{Generalisations} 
Now all this material where $a,b,c$ are real numbers can be extended to the cases in which $a, b$ and $c$ are taken from any commutative ring with identity. 
Thus, if \red{$a, b$} are integers in the ring $\Z$, then one has the discrete Heisenberg group H$_1(\Z)$.
If one takes \red{$a, b$} in $\Z_p=\Z/p\Z$ for an odd prime $p$, then one has the Heisenberg group modulo $p$, H$_1(\Z_p)$. It is a group of order $p^3$. The Heisenberg group H$_3(\Z_2)$ is of order 8 and is isomorphic to the dihedral group D$_4$ (the symmetry group of a square).
Other possibilities exist: the semi-discrete case, for which $a\in Z$ whereas $b\in [0,2\pi) \ \mathrm{mod} \ 2\pi \sim\mathbb{S}^1$, the $2$-dimensional torus $\mathbb{T}^2\sim \mathbb{S}^1\times \mathbb{S}^1$, and their higher-dimensional extensions, the $p$-adic case (see e.g. \cite{ruelle89,zelenov23} and references therein). 

The presence of the symplectic form $\mathbf{v}\wedge\mathbf{v}^{\prime}$ in \eqref{WHintL1} allows for the extension of the notion of the Heisenberg group to symplectic vector spaces. Let $(V, \boldsymbol{\omega})$ be a finite-dimensional real symplectic vector space, where $\boldsymbol{\omega}$ \red{designates}   a nondegenerate skew symmetric bilinear form on the real vector space $V$. The Heisenberg group H$(V)$ on $(V, \boldsymbol{\omega})$  is the set $V\times \R= \{(c,\mathbf{ v})\}$ endowed with the group law that mimics  Eq. \eqref{WHintL1}:
\begin{equation}
\label{symplecH}
(c,\mathbf{v})\,(c^{\prime}, \mathbf{v}^{\prime}]= \left(c+ c^{\prime} +  \frac{1}{2}\boldsymbol{\omega}(\mathbf{v},\mathbf{v}^{\prime}), \mathbf{v} + \mathbf{v}^{\prime}\right)\,.
\end{equation}
\subsection{An extra generalisation}
Another way to generalise the matrix expression \eqref{MatTrR2} consists in keeping the sub-diagonal form as:
 \begin{equation}
\label{MatTrRnsd}
\begin{split}
\R^n\in \mathbf{x}&= (x_1,x_2, \dotsb, x_n)  \leftrightarrow M_{\mathbf{x}}=
\begin{pmatrix}
    1   &   x_1       & 0         & \cdots & 0  \\
0 &   1       & x_2        & \cdots & 0   \\
\vdots  & \vdots  &\ddots    & \ddots &\vdots    \\
0 & 0  & \cdots  & 1&  x_n  \\
0 & 0  & \cdots  &  0& 1  
  \end{pmatrix} \\
  &= \bu_{n+1} + \sum_{i=1}^n x_i E_{i\; i+1}\in \mathrm{M}(n+1,\R)\, ,
  \end{split}
\end{equation}
with the notations \eqref{Eij}.
These matrices obey the multiplication and commutation rules:
\begin{equation}
\label{MMp}
M_{\mathbf{x}}\,M_{\mathbf{x}^{\prime}}= M_{\mathbf{x} + \mathbf{x}^{\prime}} + \sum_{i=1}^{n-1} x_i x^{\prime}_{i+1}E_{i \;i+2}\, , \quad \left[M_{\mathbf{x}},M_{\mathbf{x}^{\prime}}\right]=  \sum_{i=1}^{n-1} (x_i x^{\prime}_{i+1}- x_{i+1}x^{\prime}_{i})E_{i \;i+2}\,. 
\end{equation}

All these matrices, when multiplied, generate the  group SL$_{>}(n+1, \mathbb{R})$ of upper triangular matrices with all diagonal elements equal to $1$, also known as unit upper triangular matrices of order $n+1$. Although this nilpotent group is not a central extension of the translation group $\mathrm{T}_{\R^n}$, its structure is interesting to examine in light of the above Weyl-Heisenberg (WH) material.
More precisely, let us examine its structure in the manageable case $n=3$.   The 6-dimensional Lie algebra  $\mathfrak{sl}_{>}(4,\R)$ is generated by the projectors $E_{12},E_{23},E_{34},E_{13},E_{24},E_{14}$. In Tables \ref{MC3} and \ref{MCC3}   we give the multiplication and commutation rules  which  are calculated by using the formula    $E_{ij}\,E_{kl}= \delta_{jk}\, E_{il}$:
\begin{table}[H]
 \centering 
  \begin{tabular}{|c|c|c|c|c|c|c|}
\hline
$X\backslash Y$& $E_{12}$ & $E_{23}$ & $E_{34}$ & $E_{13}$ & $E_{24}$ & $E_{14}$   \\
 \hline
 $ E_{12}$  & 0&$E_{13}$  & 0 &0 & $E_{14}$ & 0\\
 \hline
 $E_{23} $& 0 & 0 &$E_{24}$ & 0& 0& 0 \\
 \hline
 $E_{34}$& 0 & 0 & 0& 0& 0& 0\\
 \hline
$ E_{13}$&0 & 0 & $E_{14}$ &0 & 0&0\\
\hline
$E_{24}$&0& 0 & 0& 0 & 0& 0\\
\hline
$ E_{14}$&0 & 0 &0 & 0&0 &0\\
\hline
\end{tabular}
 \caption{Table of matrix products $XY$}\label{MC3}
\end{table}
\begin{table}[H]
 \centering 
  \begin{tabular}{|c|c|c|c|c|c|c|}
\hline
$X\backslash Y$& $E_{12}$ & $E_{23}$ & $E_{34}$ & $E_{13}$ & $E_{24}$ & $E_{14}$   \\
 \hline
 $ E_{12}$  & 0&$E_{13}$  & 0 &0 & $E_{14}$ & 0\\
 \hline
 $E_{23} $& $- E_{13}$ & 0 &$E_{24}$ & 0& 0& 0 \\
 \hline
 $E_{34}$& 0 & $-E_{24}$ & 0& $-E_{14}$ & 0& 0\\
 \hline
$ E_{13}$&0 & 0 & $E_{14}$ &0 & 0&0\\
\hline
$E_{24}$&$-E_{14}$& 0 & 0& 0 & 0& 0\\
\hline
$ E_{14}$&0 & 0 &0 & 0&0 &0\\
\hline
\end{tabular}
 \caption{Table of commutators $[X,Y]$}\label{MCC3}
\end{table}
Denoting $V_3$, $V_2$, and $V_1$  the respective linear spans of  sets $\{E_{12},E_{23},E_{34}\}$, $\{E_{13},E_{24}\}$, and $\{E_{14}\}$, one notices the inclusion:
\begin{equation}
\label{inclusV123}
\left[V_3,V_3\right] \subset V_2\oplus V_1\, , \quad \left[V_3,V_2\right]\subset V_1\, , \quad \left[V_2\oplus V_1,V_2\oplus V_1\right]= \{0\} \,  .  
\end{equation}
Hence, we observe that $V_1$ is the center of $\mathfrak{sl}_{>}(4,\R)$, and that  $V_2\oplus V_1$ is an  abelian subalgebra. From this we infer the nilpotence of $\mathfrak{sl}_{>}(4,\R) = V_3 \oplus V_2\oplus V_1$:
\begin{equation*}
\left[\mathfrak{sl}_{>}(4,\R), \mathfrak{sl}_{>}(4,\R)\right] \subset V_2\oplus V_1\, \quad \left[\mathfrak{sl}_{>}(4,\R), V_2\oplus V_1\right] \subset V_1\, \quad \left[\mathfrak{sl}_{>}(4,\R),V_1\right]= \{0\} \,  . 
\end{equation*}
These features lead us to generalise  the WH group \eqref{WHintL1} as
the matrix in $\mathrm{SL}_{>}(4, \mathbb{R})$:
\begin{equation}
\label{GenWH3}
\begin{split}
(z,\mathbf{y},\mathbf{x}):=&  M_{\mathbf{x}} + \left(y_1+ \frac{1}{2}x_1 x_2\right)E_{13} + \left(y_2 + \frac{1}{2}x_2 x_3 \right)E_{24} \\ &+ \left(z+ \frac{1}{2}\left(x_1y_2 + x_3y_1 + x_1 x_2 x_3\right)\right)  E_{14}\\  
=& \begin{pmatrix}
   1   &  x_1 &y_1 + \frac{1}{2}x_1x_2&z + \frac{1}{2}\left(x_1y_2 + x_3y_1 + x_1 x_2 x_3\right)\\
   0   &  1 &x_2 &y_2 + \frac{1}{2}x_2x_3\\
   0   & 0 & 1 & x_3\\
  0   & 0 & 0 & 1 
\end{pmatrix}\,. 
\end{split} 
\end{equation}
The internal law is given by:
\begin{equation}
\label{intGenWH3}
\begin{split}
&(z,\mathbf{y},\mathbf{ x})(z^{\prime},\mathbf{y}^{\prime},\mathbf{ x}^{\prime})\\&= \left(z + z^{\prime} + \frac{1}{2}\Omega\left(\mathbf{x}, \mathbf{x}^{\prime},\mathbf{y},  \mathbf{y}^{\prime}\right),\mathbf{y} + \mathbf{y}^{\prime}+ \frac{1}{2}\mathsf{S}(\mathbf{x}\times\mathbf{x}^{\prime}),\mathbf{x}+ \mathbf{x}^{\prime}\right)\, .
\end{split} 
\end{equation}
where
\begin{equation}
\label{Om}
\Omega\left(\mathbf{x}, \mathbf{x}^{\prime},\mathbf{y},  \mathbf{y}^{\prime}\right)= x_1y^{\prime}_2- y_2x^{\prime}_1 + y_1x^{\prime}_3 -x_3y^{\prime}_1 -\left(x_1x^{\prime}_2x_3 + x^{\prime}_1x_2x_3+x^{\prime}_1x_2x^{\prime}_3+ x^{\prime}_1x^{\prime}_2x_3\right)\, , 
\end{equation}
\begin{equation}
 \label{om} \mathsf{S}(\mathbf{x}\times\mathbf{x}^{\prime})=\begin{pmatrix}
    0  & 0 & 1   \\
    1  &  0 & 0
\end{pmatrix} \mathbf{x}\times\mathbf{x}^{\prime}= \left\lbrace\begin{array}{c}
   x_1x^{\prime}_2- x_2 x^{\prime}_1       \\
   x_2x^{\prime}_3- x_3 x^{\prime}_2      
\end{array}\right.\,. 
\end{equation}
The unity is $e= (0,0,0,0)$ and the inverse is given by
\begin{equation}
\label{invWH1}
(z,\mathbf{y},\mathbf{ x})^{-1}= (-z,-\mathbf{y},-\mathbf{ x})\,. 
\end{equation}
The  formula \eqref{intGenWH3} issued from the parametrisation  \eqref{GenWH3} displays interesting ``subsymplectic'' structures involving two of the components of $\mathbf{x}\times \mathbf{x}^{\prime}$ and a part of the two-form $\boldsymbol{\omega}(\mathbf{v},\mathbf{v}^{\prime})$ where $\mathbf{v}:= (\mathbf{x},\mathbf{y})\in \R^5$. 
To conclude this review, we highlight another avenue for exploration: the unitary dual of the group SL$_{>}(n+1, \mathbb{R})$ (see \cite{kirillov62,kirillov04,lipsman74}) and the parametrisation of its elements of the form $z = \mathbf{x}^{(1)}, \mathbf{x}^{(2)}, \dotsc, \mathbf{x}^{(n-1)}, \mathbf{x}^{(n)}$, which generalises \eqref{GenWH3}. This direction warrants the implementation of the covariant (de-)quantisation integration program described in this paper, with potential applications to signal analysis and quantum physics.

\section{The Weyl Heisenberg group as the mathematical foundation of  time-frequency analyses}
\label{signal}
\begin{quote}
\textit{Score: a written representation of a musical composition showing all the vocal and instrumental parts arranged one below the other (Oxford English Dictionary)}
\end{quote}

\subsection{Time-frequency analysis of signals}

The two essential components of the time-frequency, or windowed Fourier transform, also known as Gabor analysis \cite{gabor46}, of a (finite-energy) temporal signal  $s(t) \in  L^2(\R) $  are translation and modulation.

\begin{figure}[H]
\begin{center}
\includegraphics[angle=0,width=12cm]{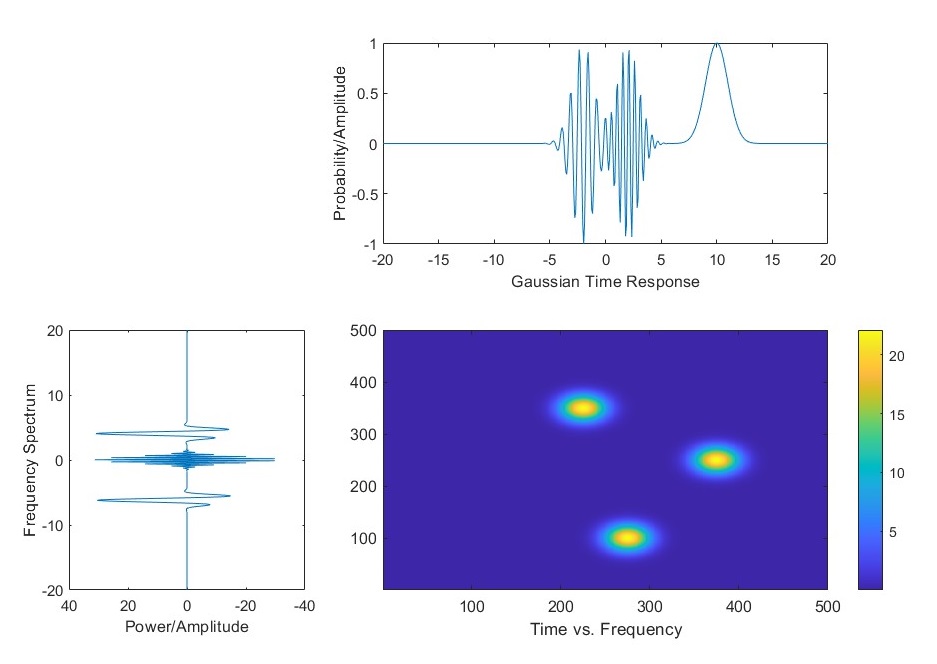}\\
\caption{Presented here is an illustrative instance of the Gabor transform, resembling a time-frequency portrait, applied to the temporal signal depicted at the top. The visualization showcases the modulus of the transform, with color gradation from low values (depicted in blue) to high values (depicted in yellow), spanning the time-frequency inverse half-plane. A comparison is drawn with the modulus of the Fourier transform of the same signal, which is concurrently displayed on the left.
}
\label{optics}
\end{center}
\end{figure}

 One chooses a \textit{probe} or \textit{window} or \textit{Gaboret} $\psi(t)\in L^2(\R,\ud t)$ which is well localized in time  and in frequency at once, and which is normalized, $\Vert \psi \Vert_{L^2(\R)} = 1$. 
 The probe is then translated in time and modulated in frequency, 
but its size is not modified (in modulus):
\begin{align}
\psi(t) \rightarrow  \psi_{ \omega,b}(t) = 
 e^{\ii\omega t}\,\psi (t -b) 
\end{align}
The time-frequency or \emph{Gabor} or \emph{windowed Fourier} transform of the signal $s(t)$ is then defined as the Hilbertian projection of the signal on the translated-modulated version of the probe:
\begin{equation} 
s(t) \rightarrow S[s](\omega , b) \equiv S(\omega,b ) = \langle \psi_{\omega,b} | s \rangle = 
\int_{-\infty}^{+ \infty}
e^{-\ii\omega t}\,\overline{\psi (t -b)} \, s(t) \, \ud t\,.
\end{equation}
An  example of a Gabor (versus Fourier) transform of a signal is shown in Fig. \ref{optics}.

The key feature of this time-frequency portrait of the signal is the conservation of its norm, \textit{i.e.},  its  energy,  which is defined as the square of the norm:
\begin{equation}
\Vert s \Vert^2_{L^2(\R)} = \int_{-\infty}^{+ \infty} \vert s(t) \vert^2 \, \ud t = 
\int_{-\infty}^{+ \infty}\int_{-\infty}^{+ \infty} \vert S(
\omega,b) \vert^2 \,\frac{\ud \omega \, \ud b }{2 \pi}\deq \Vert S \Vert^2_{L^2(\R^2)}\, .
\end{equation}
Actually this  results from the resolution of the identity fulfilled by the continuous non-orthogonal basis made of the set  $\{\psi_{ \omega, b}\, ,\, (\omega, b)\in \R^2 \}$ of translated-modulated versions of the probe:
\begin{equation}
\label{resun}
\bu = \int_{-\infty}^{+ \infty}\int_{-\infty}^{+ \infty}\frac{\ud \omega \, \ud b }{2 \pi} \left|
\psi_{\omega, b}\right\rangle \left\langle \psi_{\omega, b} \right| \,.
\end{equation}
This means that for any $\phi$ in $L^2(\R)$ we have:
\begin{equation}
\label{resunZSpp}
\Vert\phi\Vert^2= \int_{-\infty}^{+ \infty}\int_{-\infty}^{+ \infty}\frac{\ud \omega \, \ud b }{2 \pi} \vert \lg\phi|
\psi_{\omega, b}\rangle\vert^2\,. 
\end{equation}

Its consequence is the reciprocity (or \textit{reconstruction}) formula:
\begin{equation}
 s(t)= 
\int_{-\infty}^{+ \infty}\int_{-\infty}^{+ \infty}  S(\omega,b)  
e^{i\omega t}\psi (t -b)\, \frac{\ud \omega \, \ud b}{2 \pi}\,.
\end{equation}

\subsection{Weyl-Heisenberg  group and its unitary representation(s) for time-frequency analysis}
In accordance with the above notations, in time-frequency analysis the elements \eqref{defWH} of the Weyl-Heisenberg  group H$_1$ are denoted by $(c,\omega, b)$. 

In this context, the occurence of this group is understood through one of  its unitary irreducible representations (UIR). The latter can be viewed as   projective UIR's of the abelian translation group of the time-frequency plane.  Any infinite-dimensional UIR, $U^{\lambda}$, of H$_1$ is characterized by a real number $\lambda \neq 0$, and they are all equivalent
(there is also the degenerate, one-dimensional, UIR corresponding to $\lambda = 0$).
If the  Hilbert space carrying the UIR is the space  \red{$\left\{s(t) \in \mathcal{H} = L^{2}({\mathbb R} , \ud x)\right\}$}, the representation operators  are defined by the action 
\begin{equation}
\label{WHgrouprep2}
\left(U^{\lambda}(c,\omega , b)s\right)(t) = e^{\ii \lambda c}
    e^{\ii \lambda \omega(t- b/2)} s (t-b)\, . 
 \end{equation}
 The three infinitesimal generators of $U^{\lambda}$  read as:
\begin{equation} 
\begin{split}
\mathsf D^{\lambda}_c&=\ii \lambda \bu\, , \quad  (\mathsf D^{\lambda}_\omega\,s)(t) =  \ii \lambda t\, s (t)\equiv \ii \lambda \Theta s(t) \, , \quad  (\mathsf D^{\lambda}_b\, s)(t) = 
   -\frac{\partial s}{\partial t}(t) =-\ii\lambda \Omega\, s(t)\, \\
&\qquad  \qquad \qquad  \qquad \qquad   [\mathsf D^{\lambda}_\omega,\mathsf D^{\lambda}_b] = \ii \lambda\bu=  \lambda^2[\Theta,\Omega]\,.
\end{split}
\end{equation}
 Time operator $\Theta$ and frequency operator $\Omega$ are (essentially) \emph{self-adjoint}  in the Hilbert space $\mathcal{H}$ and obey the so-called canonical commutation rule (CCR) $[\Theta,\Omega]= \ii \bu/\lambda$, with its  uncertainty dispersion consequence:
 \begin{equation}
\label{uncert}
\Delta_s \Theta\, \Delta_s \Omega\geq \frac{1}{2}\, , \quad \Delta_s A:= \sqrt{\lg s| A^2|s\rg  - (\lg s|A|s\rg)^2}\,. 
\end{equation}
One should keep in mind that the CCR  $[A,B] = \ii \bu$ holds true with self-adjoint $A$, $B$,  only if both have continuous spectrum $(-\infty,+\infty)$.

 In Signal Analysis we choose $\lambda=1$  (it would be $1/\hbar$ in 1d Quantum Mechanics, with $\omega\leftrightarrow p$ and $b\leftrightarrow q$), and we simply write $U^1\equiv U$.
Hence, the representation reads as the action of exponential operators
\begin{equation}
\label{UIRWH}
\begin{split}
U(c,\omega,b)&= e^{\ii c}
    e^{-\ii \omega b/2} \, e^{\ii \omega \Theta}\, e^{-\ii b \Omega} = e^{\ii c}
    e^{\ii \omega b/2} \,e^{-\ii b \Omega} \, e^{\ii \omega \Theta}\\ 
    &\equiv e^{\ii c}\, \mathcal{D}(\omega,b)\, . 
\end{split}
\end{equation}
The expression 
\begin{equation}
\label{displW}
\mathcal{D}(\omega,b)= e^{\ii(\omega \Theta - b\Omega)}
\end{equation} 
is the unitary ``displacement or Weyl operator'', which plays a key role in the sequel. 
Let us now reinterpret the Gabor transform in terms of  Weyl-Heisenberg covariance. 
From the H$_1$ UIR action on signals:
\begin{equation}
\label{WHgrouprep2} 
(U(c,\omega,b) s )(t) = e^{\ii c}
    e^{\ii \omega(t- b/2)} s (t-b)=  e^{\ii c} (\mathcal{D}(\omega,b))s(t)\, , \quad  
    s(t) \in L^{2}({\mathbb R} \, , \ud t)\, , 
\end{equation}
 we first infer that the translation-modulation of the probe $\psi$ is a particular case of this action 
\begin{equation}
\label{partGab}
\psi(t) \rightarrow  \psi_{ \omega, b}(t)  = e^{\ii\omega t}\,\psi (t -b) = \left(U\left(\frac{\omega\,b}{2},  \omega , b\right) \psi \right)(t)\,. 
\end{equation}
The function  $ \psi_{ \omega, b}(t) $ can be viewed as a coherent state (CS) for the WH group $\mathrm{H}_1$ in the sense given by  Perelomov \cite{perel86}, and the unit-norm  $\psi$ is then called a fiducial vector. 

From the unitarity of $U$ and the representation property
 \begin{equation}
  \begin{split}
 U(c_1,\omega_1, b_1)\,U(c_2,\omega_2, b_2)&= U\left(c_1 + c_2 + \frac 1{2} (\omega_{1}b_{2} - \omega_2b_{1}), \omega_1 + \omega_2, b_1 + b_2\right)\\
&= e^{\ii (\omega_{1}b_{2} - \omega_2b_{1})} U(c_2, \omega_2, b_2)\,U(c_1,\omega_1, b_1)\, , 
\end{split}
\end{equation}
we derive the covariance of the Gabor transform
 \begin{align}
 \label{covgab}
\nonumber S[U(0,\omega_0, b_0)s](\omega, b ] &= \langle \psi_{\omega, b} | U(0,\omega_0, b_0)s \rangle
=  \langle U[0,-\omega_0, -b_0)\,U\left(\frac{ \omega\,b}{2},\omega, b\right)\psi | s \rangle\\
\nonumber &= \left\langle U\left( \frac{\omega\,b}{2} -\frac 1{2} ( \omega_{0}b - \omega b_0),\omega -\omega_0, b -b_0\right)\psi \right |\left. s \right\rangle\\
&= e^{\ii (\omega-\omega_{0}/2)b_0 }\, S(\omega -\omega_0, b -b_0)\,.
\end{align}

\section{ Weyl-Heisenberg group for the semi-discrete cylinder}
\label{sdcyl}
In this section we give a short account of the Weyl-Heisenberg symmetry and its representation when the Euclidean plane is replaced by the semi-discrete cylinder. More details are given in \cite{CIQ21}.  For other cases, see for instance \cite{feichtinger09}.

\subsection{Description of the group and its UIR}
The Weyl-Heisenberg group $\mathrm{H}^{\mathrm{dc}}_1= \{(s, m,\theta)\in \R\times\Z\times \mathbb{S}^1\}$  for the semi-discrete cylinder $\Gamma=\Z\times \mathbb{S}^1$ is equipped with the internal law 
\begin{equation}
\label{smintern}
(s, m,\theta)(s^{\prime}, m^{\prime},\theta^{\prime})= \left(s+s^{\prime} +\frac{m\theta^{\prime}-m^{\prime}\theta}{2},m+m^{\prime},\theta+\theta^{\prime} \, \mathrm{mod}\,2\pi\right)\,. 
\end{equation}
This group  is the discrete \& compact  version  of the Weyl-Heisenberg group H$_1$. 
We now view $\Gamma$ as the left coset $\mathrm{H}^{\mathrm{dc}}_1/\R$. For any element $(m,\theta)$ of this coset and for any circular signal  in $L^2(\mathbb{S}^1,\ud\gamma)$ one defines the  unitary ($\sim$ Weyl or displacement) operator $\mathcal{D}$ 
as:
\begin{equation}
\label{UIR}
(\mathcal{D}(m,\theta)\phi)(\gamma)=e^{-\ii\,\frac{m\theta}{2}}\,e^{\ii\,m\gamma}\phi(\gamma-\theta)\,.
\end{equation}
Note a few useful formulas: 
\begin{equation}
\label{UUp}
\mathcal{D}(m,\theta)\,\mathcal{D}(m^{\prime},\theta^{\prime})= e^{\ii\frac{m\theta^{\prime}-m^{\prime}\theta} {2}}\,\mathcal{D}(m+m^{\prime},\theta+\theta^{\prime})\,. 
\end{equation}
\begin{equation}
\label{UUUp}
\mathcal{D}(m,\theta)\mathcal{D}(m^{\prime},\theta^{\prime})\,\mathcal{D}^{\dag}(m,\theta)= e^{\ii(m\theta^{\prime}-m^{\prime}\theta)}\mathcal{D}(m^{\prime},\theta^{\prime})\,. 
\end{equation}
\begin{equation}
\label{TrU}
\mathrm{Tr}[\mathcal{D}(m,\theta)]= \sum_{n\in \Z} \lg e_n|\mathcal{D}(m,\theta)|e_n\rg= 2\pi\delta_{m\,0}\delta(\theta)\,, 
\end{equation}
where the functions $e_n(\gamma)= \frac{1}{\sqrt{2\pi}}e^{\ii n \gamma}$ form the orthonormal Fourier basis of $L^2(\mathbb{S}^1,\ud\gamma)$. Formula \eqref{TrU} is established thanks to the Poisson summation formula restricted to $\mathbb{S}^1$:
\begin{equation}
\label{poissonS1}
 \frac{1}{2\pi}\sum_{n\in \Z} e^{\ii n t} = \sum_{n\in \Z} \delta(t+ 2\pi n) = \delta(t) \quad \mbox{for}\quad t \in [0,2\pi)\,, 
\end{equation}
and the expression of the matrix elements  of the displacement operator with respect to the Fourier basis
\begin{equation}
\label{mateleD}
\lg e_n|\mathcal{D}(m,\theta)|e_n\rg=\left(\mathcal{D}(m,\theta)\right)_{nn^{\prime}}= e^{\ii (\frac{m}{2}-n)\theta}\, \delta_{n-m\,n^{\prime}}\,. 
\end{equation}
\subsection{Resolutions of the unity by rotated-\&-discretely modulated probes}
Akin the continous case, one shows (e.g., see \cite{CIQ21}) that for any $\psi$ in $L^2(\mathbb{S}^1,\ud\gamma)$, $\Vert\psi\Vert=1$,  the family $\{ \psi_{m,\theta}\equiv \mathcal{D}(m,\theta)\psi\}_{(m,\theta)\in\Gamma}$ constitutes an overcomplete  family  resolving the identity :
\begin{equation}
\label{resunZS}
\frac{1}{2\pi} \sum_{m\in\Z}\int_{\mathbb{S}^1}\ud\theta\, | \psi_{m,\theta}\rangle\langle \psi_{m,\theta}| = \mathbbm{1}\, .
\end{equation}
 This implies that for any $\phi$ in $L^2(\mathbb{S}^1,\ud\gamma)$ we have:
\begin{equation}
\label{resunZSpp}
\frac{1}{2\pi}  \sum_{m\in\Z}\int_{\mathbb{S}^1}\ud\theta\,\vert{\langle  \psi_{m,\theta}|\phi\rangle\vert}^2= \Vert\phi\Vert^2 \,. 
\end{equation}
Similarly to the continuous case, the fiducial vector  $ \psi_{m,\theta}:=\mathcal{D}(m,\theta)\psi$ can be viewed as a Perelomov coherent state  for the symmetry group $\mathrm{H}^{\mathrm{dc}}_1$. The interpretation of the projection  of $\phi$  on $\psi_{(m,\theta)}$, namely
\begin{equation}
\label{scport}
\langle\,\psi_{m,\theta}\vert\phi\rangle
=\int_{\mathbb{S}^1}\ud\gamma\, \overline{\psi_{m,\theta}(\gamma)}\phi(\gamma)
= \int_{\mathbb{S}^1}\ud\gamma\, e^{\ii\frac{m\theta}{2}}e^{-\ii\,m\gamma}\overline{\psi(\gamma-\theta)}\phi(\gamma)\, , 
\end{equation}
is the  angular momentum - angle  (or phase space) representation of $\psi$ with respect to the family of  coherent states $\psi_{m,\theta}$.

Within the context of analysis of signals living on the unit circle, Eq.\,\eqref{scport} is the Gabor or windowed Fourier transform of periodic signals, except that, in that context, it is  usual to drop the phase $e^{-\ii \frac{m\theta}{2}}$. Then, for a unit-norm  fiducial vector $\psi\in L^2(\mathbb{S}^1)$ and a $2\pi$-periodic signal $\phi\in  L^2(\mathbb{S}^1)$ one can also adopt the notation for its Gabor transform:
\begin{equation}
\label{Gabtrans}
S^{\psi}[\phi](m,\theta)= \int_{\mathbb{S}^1} \ud \gamma \, \overline{\psi_{m,\theta}(\gamma)}\,\phi(\gamma)= \lg \psi_{m,\theta}|\phi\rg\,,
\end{equation}
and, consistently to \eqref{resunZS},  the Gabor reconstruction of signal $\phi$ reads as
\begin{equation}
\label{reconsgab}
\phi(\gamma)= \frac{1}{2\pi}  \sum_{m\in\Z}\int_{\mathbb{S}^1}\ud\theta \,S^{\psi}[\phi](m,\theta)\, \psi_{m,\theta}(\gamma)\,. 
\end{equation}
In the case where  the fiducial vector is a coherent state $\psi_{m^{\prime},\theta^{\prime}}(\gamma)$, its Gabor transform $\lg \psi_{m,\theta}|\psi_{m^{\prime},\theta^{\prime}}\rg$ is a  reproducing kernel for the Hilbert space of square integrable functions on $\Gamma$ equipped with  the scalar  product $\lg \Phi | \Phi^{\prime}\rg = \frac{1}{2\pi}  \sum_{m\in\Z}\int_{\mathbb{S}^1}\ud\theta \, \overline{\Phi(m,\theta)}\,\Phi^{\prime}(m,\theta)$.

\begin{figure}[H]
    \centering
    \includegraphics[angle=0,width=12.5cm]{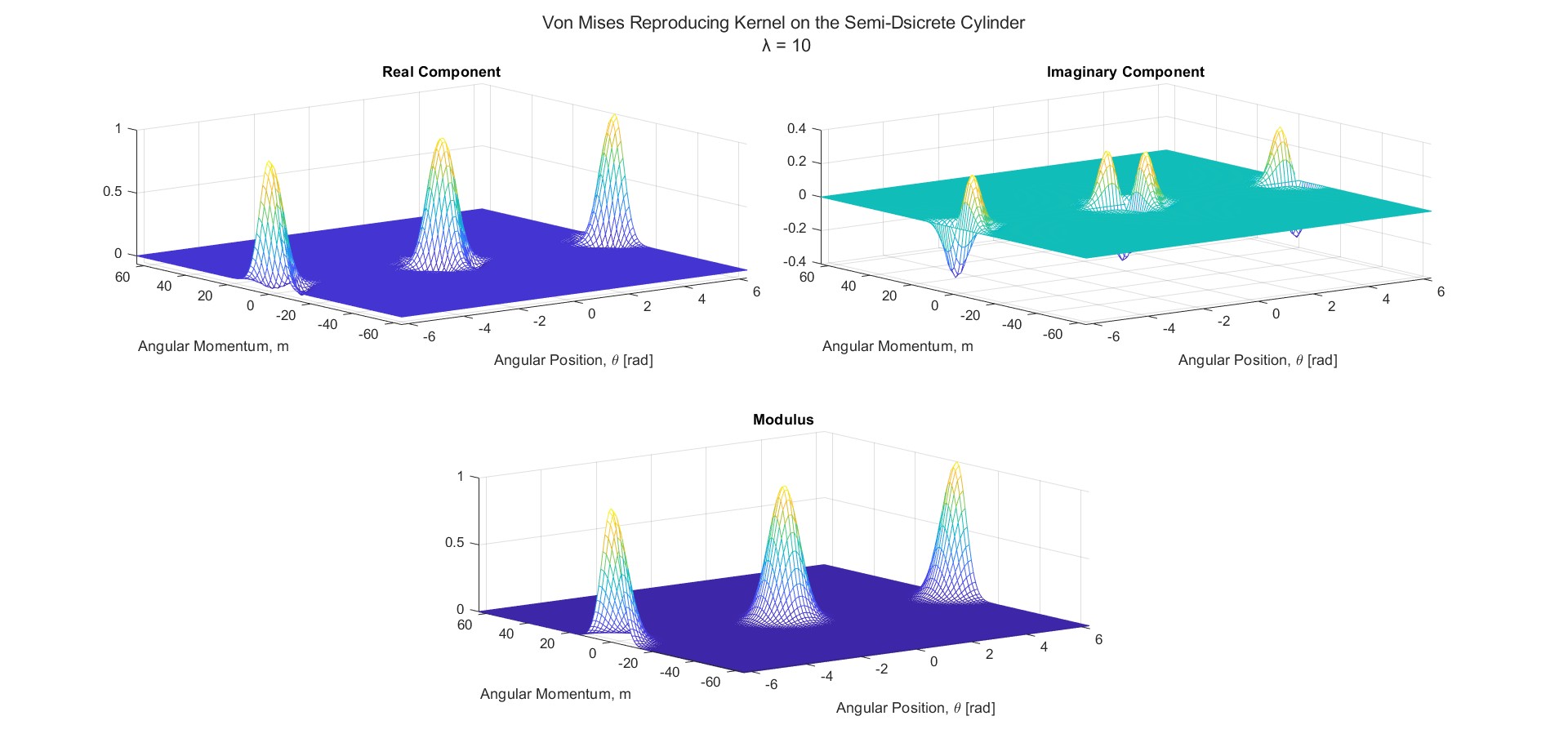}
    \caption{This illustration shows the real component (top right), the imaginary component(top left), and the modulus(bottom) of the \red{von} Mises reproducing kernel. As this is representative of the semi-discrete cylinder, multiple periods are shown with each plot being defined $\in [-2\pi, 2\pi]$.}
    \label{VonMises Fig}
\end{figure}

In Figure \ref{VonMises Fig}, we illustrate this formalism with the so-called \red{von} Mises fiducial vector
\begin{equation}
    \psi(\gamma)=\frac{e^{\lambda\cos\gamma}}{\sqrt{2\pi\,I_{0}(2\lambda)}}\,. \label{VonMises} 
\end{equation}
The corresponding reproducing kernel reads:
\begin{equation}
    \lg\psi_{m,\theta}|\psi_{m', \theta'}\rg = \frac{e^{\ii\frac{m\theta^{\prime} - m^{\prime}\theta}{2}}} {2\pi\,I_0(2\lambda)}\,I_{m-m'}\left(2\lambda \cos\left(\frac{\theta - \theta'}{2}\right)\right)\, , 
\end{equation}
where $I_\nu$ is the modified Bessel of the first kind of zero order. 

\section{Playing with quantization and density operators: general}
\label{quantiz}

\subsection{Integral quantization}
\label{iquantiz}
The integral quantization of a function $f(x)$ on a measure space $(X,\nu)$ is defined by the linear map 
\begin{equation}
\label{iqmap}
	f\mapsto A^{\sf M}_{f}=\int_{X}\,{\sf M}(x)\, f(x)\,\ud\nu(x)\,,
\end{equation}
where the family of  matrices ${\sf M}(x)$  solves (in a weak sense) the identity  as
\begin{equation}
\label{resunitM}
	X\ni x\mapsto\mathsf{M}(x)\,,\quad\int_{X}\,{\sf M}(x)\,\ud\nu(x)=\bu\,.
\end{equation} 

If the matrices ${\sf M}(x)$ are non-negative, the quantum operator related to the characteristic function on $\Delta$, $A^{\sf M}_{\chi_\Delta}$, defines a POVM . Indeed, the restriction to $\Delta$ of the quantization map \eqref{iqmap} 
\begin{align}
	F(\Delta):=A^{\sf M}_{\chi_\Delta}=\int_{X}\,{\sf M}(x)\, \chi_\Delta(x)\,\ud\nu(x)=\int_{\Delta}\,{\sf M}(x)\, \ud\nu(x) \;,
\end{align}
solves the identity  by definition. Therefore,  POVMs can play  two key roles: they are the mathematical representatives of observables and they provide a quantization procedure.

Let us suppose that the operators ${\sf M}(x)$ are unit trace and non-negative, \textit{i.e.}, are  density operators,  ${\sf M}(x)=\rho_{\mathrm{a}}(x)$ (subscript `a' is for ``analysis''). Then, the quantisation \eqref{iqmap} maps any real probability distribution $\mathfrak{w}$ on $(X,\nu)$, \textit{i.e.}, non-negative function $\mathfrak{w}(x)$, with unit average on $(X,\nu)$:
 \begin{equation}
\label{avf}
\lg \mathfrak{w} \rg_{X,\nu}= \int_{X}\,\, \mathfrak{w}(x)\,\ud\nu(x)= 1\, ,
\end{equation}
to the density operator
\begin{equation}
\label{iqmapvp}
	\mathfrak{w} \mapsto \rho^{\rho_{\mathrm{a}}}_{\mathfrak{w}}=\int_{X}\,\rho_{\mathrm{a}}(x)\, \mathfrak{w}(x)\,\ud\nu(x)\,. 
\end{equation}
This representation of a given mixed state as a continuous convex superposition of mixed states 
 provides one more  illustration of a \textit{typical property of quantum-mechanical ensembles in comparison with their classical counterparts.}

\subsection{Semi-classical portraits}
\label{sdport}
A quantum model for a system must have a classical or semi-classical counterpart. We say there exists a \textit{dequantization} map acting on an operator $A_f$ to give back the original $f(x)$, in general with some corrections. This procedure gives a \textit{semi-classical portrait} of the original function $f$, often denoted $\check f(x)$ (Husimi and Wigner functions,  lower (Lieb)  or covariant (Berezin) symbols). Given two families of operators, ${\sf M}_{\mathrm{a}}(x)$ (for ``analysis'') and ${\sf M}_{\mathrm{r}}(x)$ (for ``reconstruction"), resolving the identity in the sense of \eqref{resunitM},
we here generalize  the definition of these lower symbols in the following way:
\begin{equation}
 \label{lowsymbgen}
A_{f}\mapsto\check{f}^{\mathrm{M}_{\mathrm{ar}}}(x):=\mathrm{tr}(\mathsf{M}_{\mathrm{r}}(x)\, A^{\sf M_{\mathrm{a}}}_{f})= \int_{X}\,\mathrm{tr}(\mathsf{M}_{\mathrm{r}}(x){\sf M}_{\mathrm{a}}(x^{\prime}))\, f(x^{\prime})\,\ud\nu(x^{\prime})\,,
\end{equation} 

Now choosing $\mathsf{M}_{\mathrm{a}}(x)=\rho_{\mathrm{a}}(x)$ and $\mathsf{M}_{\mathrm{r}}(x)=\rho_{\mathrm{r}}(x)$ to be  density matrices, the map $x^{\prime} \mapsto \mathrm{tr}(\rho_{\mathrm{r}}(x)\rho_{\mathrm{a}}(x^{\prime}))$ defines a probability
distribution $\mathrm{tr}(\rho_{\mathrm{r}}(x)\rho_{\mathrm{a}}(x^{\prime}))$ on the measure space $(X,\ud\nu(x^{\prime}))$, as it is easily proved by multiplying the resolution of the unity \eqref{resunitM} with $\rho_{\mathrm{r}}(x)$ and tracing the result. Thus, the expectation value $\check{f}^{\rho_{\mathrm{ar}}}(x)$ of the operator $A^{\rho_{\mathrm{a}}}_f$  is given by:
\begin{equation}
 \label{locaver}
f(x)\mapsto\check{f}^{\rho_{\mathrm{ar}}}(x)=\int_{X}f(x^{\prime})\,\mathrm{tr}(\rho_{\mathrm{r}}(x)\rho_{\mathrm{a}}(x^{\prime}))\,\mathrm{d}\nu(x^{\prime})\,.
\end{equation}
The map \eqref{locaver} represents in general a regularization of the original, possibly extremely singular, $f$. Standard cases are for $\rho_{\mathrm{a}}= \rho_{\mathrm{r}}$, in particular for rank-one density matrices (coherent states).
Now, if we apply this scheme to the probability distribution  case \red{where} $f=\mathfrak{w}$ is a probability distribution on $(X,\nu)$, we then obtain the interesting result:
\begin{equation}
 \label{locavervp}
\mathfrak{w}(x)\mapsto\widecheck{\mathfrak{w}}^{\rho_{\mathrm{ar}}}(x)=\int_{X}\mathfrak{w}(x^{\prime})\,\mathrm{tr}(\rho_{\mathrm{r}}(x)\rho_{\mathrm{a}}(x^{\prime}))\,\mathrm{d}\nu(x^{\prime})\,.
\end{equation}
This equality can  be viewed as a kind of convolution of two probability distributions on the measure space $(X,\nu)$.  Consider that \eqref{locavervp} can be interpreted as obtaining from a purely classical probability distribution on $(X,\nu)$ a new one with a quantum nuance encapsulated by the pair $\left(\rho_{\mathrm{a}}, \rho_{\mathrm{r}}\right)$. 
Of course, we deviate from the standard definition, which typically involves the addition of independent random variables and, by extension, to forming linear combinations of random variables.  This deviation arises because, so far, we have not precisely defined the composition law appearing in the kernel  $\mathrm{tr}(\rho_{\mathrm{r}}(x)\rho_{\mathrm{a}}(x^{\prime}))$.

\subsection{Covariant integral quantization}
\label{CIQ}
Covariance implies symmetry. Symmetry suggests a mathematical group
$G$. A symmetry structure on $X$ presupposes that some group $G$
acts on $X$. Our hypothesis here is that $X$ is an homogeneous space
for $G$, which means that $X\sim G/H$ for $H$ a subgroup of $G$.  
We now examine two possible cases. 
\subsubsection*{(i) $X=G$: quantizing the group}
Let $G$ be a Lie group with left invariant Haar measure $\ud\mu(g)$, and let
$g\mapsto U(g)$ be a UIR
of $G$ in a Hilbert space $\mathcal{H}$. Let ${\sf M}$ be a bounded
operator on $\mathcal{H}$. Suppose that the operator 
\begin{equation} \label{}
\mathsf{R}:=\int_{G}\,{\sf M}(g)\,\ud\mu(g)\,,\quad{\sf M}(g):=U(g)\,{\sf M}\, U(g)^{\dag}\,,
\end{equation} 
is defined in a weak sense. From the left invariance of $\ud\mu(g)$
the operator $\mathsf{R}$ commutes with all operators $U(g)$, $g\in G$, and so, from
Schur's Lemma, $\mathsf{R}=c_{{\sf M}}\bu$ with 
\begin{equation} \label{}
c_{{\sf M}}=\int_{G}\,\mathrm{tr}\left(\rho_{0}\,{\sf M}(g)\right)\,\ud\mu(g)\,,
\end{equation} 
where the unit trace positive operator $\rho_{0}$ is chosen in order
to make the integral convergent. If $c_{{\sf M}}$ is finite and not zero, then
we get the resolution of the identity: 
\begin{equation} \label{resunitM}
\int_{G}\,{\sf M}(g)\,\ud\nu(g)=\bu\,,\quad\ud\nu(g):=\ud\mu(g)/c_{{\sf M}}\,.
\end{equation} 
For this reason, the operator $\mathsf{M}$ to be $U$ transported will be designated (admissible) fiducial operator. 
The method becomes more apparent when the representation $U$ is
square-integrable in the following sense (for example, it is the
case for the affine group of the real line). Suppose that there exists
a density operator $\rho$ which is admissible in the sense that 
\begin{equation} \label{crho}
c_\rho=\int_{G}\ud\mu(g)\,\mathrm{tr}(\rho\,\rho(g))<\infty\,,
\end{equation} 
with $\rho(g)=U(g)\rho U^{\dag}(g)$. Then the resolution of the identity \eqref{resunitM} holds
with ${\sf M}(g)$ and $c_{{\sf M}}= c_\rho$. This allows the \textit{covariant} integral quantization of complex-valued
functions on the group 
\begin{equation}
\label{qmapintU}
f\mapsto A^{\rho}_{f}=\int_{G}\,{\rho(g)}\, f(g)\,\ud\nu(g)\,,\quad \ud\nu(g):= \frac{\ud\mu(g)}{c_\rho}
\end{equation}
\begin{equation}\label{}
U(g)A^{\rho}_{f}U^{\dag}(g)=A^{\rho}_{\mathcal{U}(g)f}\,,\quad\mbox{(covariance)}
\end{equation} 
where 
\begin{equation} \label{regrep}
(\mathcal{U}(g)f)(g'):=f(g^{-1}g')
\end{equation} 
is the regular representation if $f\in L^{2}(G,\ud\mu(g))$. This leads to a non-commutative version of the original manifold structure for $G$. 

\subsubsection*{(ii) $X=G/H$: quantizing a non-trivial group coset}
In the absence of square-integrability over $G$ (as is seen above, the trivial illustration
is the Weyl-Heisenberg group), there exists (\cite{aag14} a definition of square-integrable representations with respect to a left coset manifold $X=G/H$, with
$H$ a closed subgroup of $G$ (it is the center in the Weyl-Heisenberg case), equipped with a quasi-invariant measure
$\nu$ . For a global Borel section $\sigma:X\rightarrow G$
of the group, let $\nu_{\sigma}$ be the unique quasi-invariant measure
defined by 
\begin{equation} \label{}
{\ud}\nu_{\sigma}(x)=\lambda(\sigma(x),x)\,{\ud}\nu(x)\,,
\end{equation} 
where $\lambda(g,x){\ud}\nu(x)={\ud}\nu(g^{-1}x)\;\;(\forall g\in G)$ with $\lambda(g,x)$  obeying the cocycle condition
$\lambda(g_1 g_2,x)= \lambda(g_1,x)\,\lambda(g_2, g_1^{-1}x)$.
Let $U$ be a UIR which is square integrable $\mathrm{mod}(H)$ with 
an admissible $\rho$, \textit{i.e.}, $c_\rho:=\int_{X}\,\mathrm{tr}\left(\rho\,\rho_{\sigma}(x)\right)\,{\ud}\nu_{\sigma}(x)<\infty$, 
with $\rho_{\sigma}(x)=U(\sigma(x))\rho U(\sigma(x))^{\dag}$. Then we
have the resolution of the identity and the resulting quantization
\begin{equation}\label{}
f\mapsto A^{\rho_\sigma}_{f}=\frac{1}{c_\rho}\int_{X}\; f(x)\,\rho_{\sigma}(x)\,{\ud}\nu_{\sigma}(x)\,.
\end{equation} 
In order to establish the covariance of this quantization let us consider  the sections $\sigma_{g}: X \rightarrow G, \;\; g \in G$, which are 
covariant translates of $\sigma$ under $g$:
\begin{equation}
 \sigma_{g}(x) = g\sigma (g^{-1}x) = \sigma (x)h(g, g^{-1}x)\, ,
\end{equation}
where $h$ is the cocycle defined by 
\begin{equation}
g\sigma (x) = \sigma (gx)h(g,x) \quad \mbox{with} \quad
h(g,x)  \in H\, .
\label{cocyclecond2}
\end{equation}
With 
 ${\ud}\nu_{\sigma_g}(x) := \lambda (\sigma_{g} (x), x)\,{\ud}\nu$
 one introduces 
 \begin{equation}
\label{rhosig}
 \rho_{\sigma_g}(x) = 
U(\sigma_{g}(x))\,\rho\, U(\sigma_{g}(x))^{\dag}\, .
\end{equation}
Hence, for a UIR $U$ which is square integrable  $\mathrm{mod}(H, \sigma )$, 
a general covariance property of the integral quantization  $f\mapsto A_f$ reads as
\begin{equation}
\label{quantizcovmodH2}
 U(g) A^{\rho_{\sigma}}_f U(g)^{\dagger} = A^{\rho_{\sigma_g}}_{\mathcal{U}_l(g)f}\, , \quad A^{\sigma_g}_f:= \frac{1}{c_\rho}\int_X \rho_{\sigma_g}(x) f(x) \ud \nu_{\sigma_g}(x)\,, 
\end{equation}
where  $(\mathcal{U}_l(g)f)(x)=f\left(g^{-1}x\right)$. 
Similar results hold true by replacing $\rho$ by  more general bounded operators ${\sf M}$ provided integrability and weak convergence hold in the above expressions. 

\subsection{An example of construction of original operator $\sfM$ or $\rho$}
\label{weight}
Let $U$ be a UIR of $G$ and $w(x)$ be a function on the coset $X= G/H$.
Suppose that it allows to define a bounded operator ${\sf M}$ on $\calH$ through the operator-valued integral
\begin{equation}
\label{opMvarpi}
{\sf M}_{\sigma}^{\vap}= \int_{X} w(x)\,U(\sigma(x))\,  {\ud}\nu_{\sigma}(x)\, . 
\end{equation}
Then, under appropriate conditions on the ``weight'' function $w(\sigma(x))$ such that   $U$ be a UIR which is square integrable $\mathrm{mod}(H,\sigma)$ and ${\sf M}$ is admissible in the above sense,  the family of transported operators  ${\sf M}_{\sigma}^{\vap}(x):= U(\sigma(x)) {\sf M}_{\sigma}^{\vap}U^\dagger(\sigma(x))$ resolves the identity.  

An interesting problem concerns the reciprocal construction of the weight $w$, \textit{i.e.},  find the operator $V$ such that 
\begin{equation}
\label{invweight}
w(x) = \mathrm{tr} \left( {\sf M}_{\sigma}^{w} V(\sigma(x))\right)\, , \ \mbox{\textit{i.e.}}\, ,  \  V(\sigma(x))U(\sigma(x^{\prime})) = \delta^\sigma_\nu(x,x^{\prime})\, , 
\end{equation}
where the Dirac  measure $\delta^\sigma_\nu$ is defined as 
\begin{equation}
\label{diracnu}
\int_X {\ud}\nu_{\sigma}(x)\, \delta^\sigma_\nu(x,x^{\prime})\, f(x^{\prime})= f(x)\,.  
\end{equation}
This question generalises to the unitary dual of a group the notion of inverse Fourier transform for the group $\R$. 

\section{Weyl-Heisenberg covariant integral quantization(s)}
\label{WHintquant}

We now illustrate the above material with a few examples involving the continuous Weyl-Heisenberg group. Other examples involving other types of the Weyl-Heisenberg symmetry are presented in the published papers \cite{CIQ21,fabre23}. 

With the  notations introduced above,  the group is $\sf{H}_1$ and the subgroup $H$ is its center $C$. Thus, the measure space $(X,\ud \nu)$ is the Euclidean plane equipped with the Lebesgue measure,
\begin{equation}
\label{WHX}
X= \R^2= \{(\omega,b)\,,\, \omega,b\in \R\}\, , \quad \ud \nu(\omega,b)= \frac{\ud \omega\,\ud b}{2\pi}\,. 
\end{equation}
 The unitary operators we just need are the UIR operators modulo the phase factor $e^{\ii c}$ in \eqref{UIRWH}, \textit{i.e.}, the displacement operator $\mathcal{D}(\omega,b)$ in \eqref{displW}. The three key formulae \eqref{avf}, \eqref{iqmapvp}, and \eqref{locavervp} become in the present context
\begin{equation}
\label{WHavf}
 \int_{\R^2}\, \mathfrak{w}(\omega,b)\,\frac{\ud \omega\,\ud b}{2\pi}= 1\, ,
\end{equation}
\begin{equation}
\label{WHiqmapvp}
\rho^{\rho_{\mathrm{a}}}_{\mathfrak{w}}=\int_{\R^2}\,\rho_{\mathrm{a}}(\omega,b)\, \mathfrak{w}(\omega,b)\,\frac{\ud \omega\,\ud b}{2\pi}\,, \quad \rho_{\mathrm{a}}(\omega,b)= \mathcal{D}(\omega,b)\rho_a \mathcal{D}(-\omega,-b)\, .
\end{equation} 
\begin{equation}
 \label{WHlocavervp}
\widecheck{\mathfrak{w}}^{\rho_{\mathrm{ar}}}(\omega,b)=\int_{\R^2}\mathfrak{w}(\omega^{\prime},b^{\prime})\,\mathrm{tr}(\rho_{\mathrm{r}}(\omega,b)\rho_{\mathrm{a}}(\omega^{\prime},b^{\prime}))\,\frac{\ud \omega^{\prime}\,\ud b^{\prime}}{2\pi}\,.
\end{equation}

\subsection{Playing with Gabor probes}

Let us start with a normalised probe $\psi_{\mathrm{a}} \in L^2(\R,\ud t)$, $\Vert\psi_{\mathrm{a}}\Vert=1$, and its corresponding projector $\rho_{\mathrm{a}}= |\psi_{\mathrm{a}}\rg\lg\psi_{\mathrm{a}}|$.  We remind from \eqref{WHgrouprep2} the action of the displacement operator
\begin{equation}
\label{distprobe}
(\mathcal{D}(\omega,b)\psi_{\mathrm{a}})(t)= e^{\ii(\omega (t-b/2)}\psi_{\mathrm{a}}(t-b)=e^{-\ii\omega b/2}\psi_{\mathrm{a};\omega,b}(t)\,. 
\end{equation}
Hence, as acting on a vector $s\in L^2(\R,\ud t)$, the operator $\rho^{\psi_{\mathrm{a}}}_{\mathfrak{w}}$ read as the integral operator
\begin{equation}
\label{Rintop}
\left(\rho^{\psi_{\mathrm{a}}}_{\mathfrak{w}}s\right)(t)= \int_{-\infty}^{\infty}\mathcal{R}(t,t^{\prime}) \, s(t^{\prime})\,\ud t^{\prime}\, ,
\end{equation}
with kernel
\begin{equation}
\label{KerRintop}
\mathcal{R}(t,t^{\prime}) =\frac{1}{\sqrt{2\pi}} \int^\infty_{-\infty}\ud b\, \widehat{\mathfrak{w}}_p(t^{\prime}-t,b)\, \overline{\psi}_{\mathrm{a}}(t^{\prime}-b)\,\psi_{\mathrm{a}}(t-b)\, , 
\end{equation}
where $\widecheck{\mathfrak{w}}_p$ is the partial Fourier transform of $\mathfrak{w}$ with respect to the first variable, 
\begin{equation}
\label{pfourtr}
\widehat{\mathfrak{w}}_p(\xi,b)= \frac{1}{\sqrt{2\pi}} \int^\infty_{-\infty}\ud \omega \,e^{-\ii \omega \xi}\, \mathfrak{w}(\omega,b)\,. 
\end{equation}
We now examine the semi-classical distribution resulting from the above quantisation.

Pick a second  normalised probe $\psi_{\mathrm{r}} \in L^2(\R,\ud t)$, $\Vert\psi_{\mathrm{r}}\Vert=1$ to build the new probability density $\widecheck{\mathfrak{w}}$ :
  \begin{equation}
\label{WHlocavervp}
\widecheck{\mathfrak{w}}^{\psi_{\mathrm{ar}}}(\omega,b)=
\int_{\R^2}\mathfrak{w}(\omega^{\prime},b^{\prime})\,\left\vert \lg\psi_{\mathrm{r};\omega,b}| \psi_{\mathrm{a};\omega^{\prime},b^{\prime}}\rg\right\vert^2\,\frac{\ud \omega^{\prime}\,\ud b^{\prime}}{2\pi}
= \left(\mathfrak{w}\ast\mathsf{P}^{\mathrm{ar}}\right)(\omega,b)\, , 
\end{equation}
where the probability distribution $\mathsf{P}^{\mathrm{ar}}$ on $\left(\R^2,\frac{\ud\omega\,\ud b}{2\pi}\right)$ is built from the probes as
\begin{equation}
\label{Prob_4Gauss}
\mathsf{P}^{\mathrm{ar}}(\omega, b)= \int_{\R^2}\ud\tau\,\ud\tau^{\prime}\,e^{\ii\omega(\tau^{\prime}-\tau)}\, \overline{\psi}_{\mathrm{r}}(\tau)\,\psi_{\mathrm{r}}(\tau^{\prime})\,\psi_{\mathrm{a}}(\tau +b)\,\overline{\psi}_{\mathrm{a}}(\tau^{\prime}+b)\, . 
\end{equation}
The convolution $ \mathfrak{w}\ast\mathsf{P}^{\mathrm{ar}}$
is precisely the probability distribution  for the difference of two vectors in the time-frequency plane, viewed as independent random variables,  and fits to the abelian and homogeneous structure of the time-frequency plane (remind that the choice of its origin   is left arbitrary).
\subsection{Examples of Probability Density Distributions on the Time-Frequency Plane}
\begin{figure}[H]
    \centering
    \includegraphics[angle=0,width=6cm]{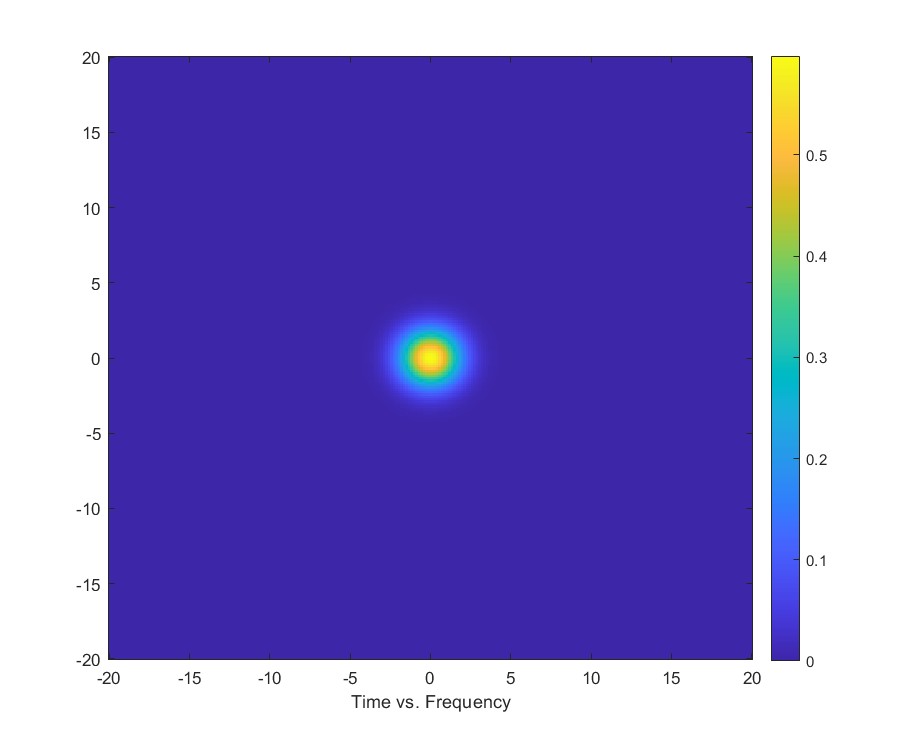}
    \includegraphics[angle=0,width=6.5cm]{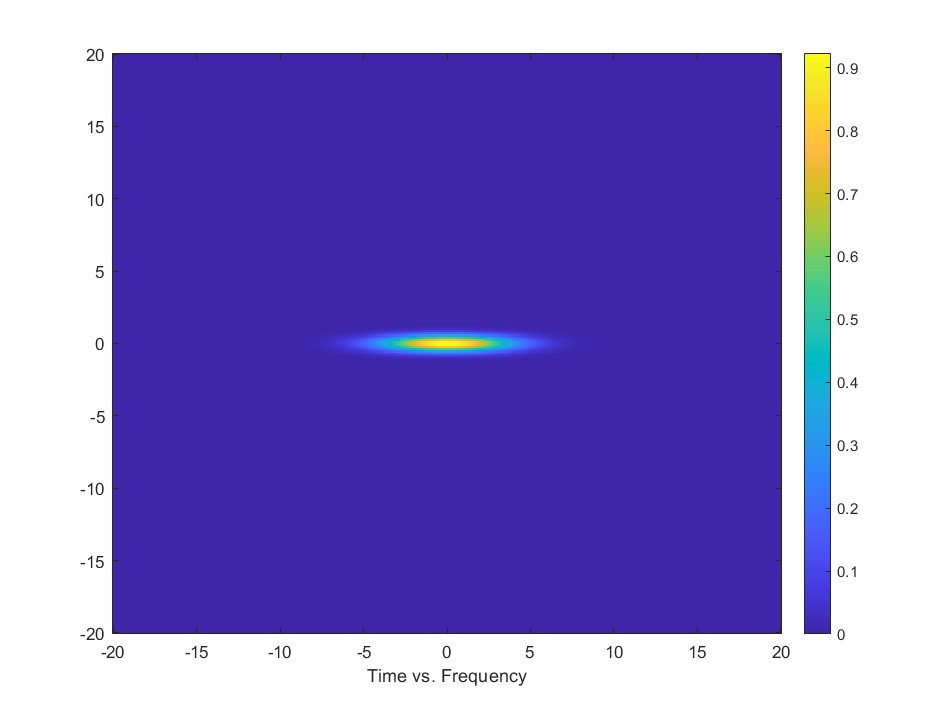}
    \caption{In this figure we observe the probability distribution $\text{P}^{ar}(\omega,b)$ acting on the time-frequency plane. The two parameters $r$ and $a$ determine the distortion of our distribution. In one special case, when $a$ is equal to the reciprocal of $r$, the resulting probability distribution takes the typical form of two dimensional gaussian. The probes used to generate this function are $L^2(\mathbb{R})$ normalized Gaussian packets. Here the color represents the amplitude of the function, \textit{i.e.}, the probability for a given input of frequency and time. \red{Parameters values are: $a=5$, $r=1/5$ (left), $a= 5$, $r=10$ (right).}}
    \label{Gauss_PhaseSpace}
\end{figure}

We consider the following examples:
\begin{equation}
\label{psia}
    \psi_a(\tau) = \frac{1}{(\pi a)^{1/4}}e^{-\frac{1}{2}\frac{\tau^2}{a}}
\end{equation}
and 
\begin{equation}
\label{psib}
    \psi_r(\tau) = \frac{1}{(\pi r)^{1/4}}e^{-\frac{1}{2}\frac{\tau^2}{r}}.
\end{equation}
When continuing equation \ref{Prob_4Gauss} using these gaussian probes, $\text{P}^{ar}$ yields
\begin{align}
    \text{P}^{ar}(\omega,b) = \frac{2\sqrt{\pi a}}{r+a}e^{-\frac{ra}{r+a}\omega^2}e^{\frac{-b^2}{r+a}}.
    \label{Prob_4Gauss_plugged}
\end{align}
Utilizing Equation \ref{Prob_4Gauss_plugged}, we are able to utilize this two dimensional gaussian on our time-frequency phase space, as shown in \red{Figure} \ref{Gauss_PhaseSpace}.
\section{Stellar Majorana}
\label{stellar}

In this section, we provide insights into the applications of our formalism to stellar Majorana representations for mixed states.

 The Majorana stellar representation \cite{majorana32} (see for instance the recent \cite{serrano20,chabaud22})  is a way to geometrically visualize pure spin-$s$ states. 
 In essence, the Majorana stellar representation for spin establishes a bijection between Hilbert space states and complex polynomials, and subsequently, a bijection between the roots of these polynomials and sets of points on the unit sphere, $\mathbb{S}^2\cong \mathrm{SU}(2)/\mathrm{U}(1)$. 
 
 We remind that $\mathbb{S}^2$, as a Kahlerian manifold,  is viewed as the set of classical spin states, \textit{i.e.} is the classical phase space for the spin. 
 
In their phase space representation the SU$(2)$ and at given $j\in \N/2$, spin coherent states   $|\theta,\phi\rangle\equiv |\widehat{\mathbf{n}}\rg\in \C^{2j+1}$  \cite{perel86} read as the overlap of this CS with  another CS, $|\eta,\psi\rangle\equiv |\widehat{\mathbf{r}}\rg$:
\begin{equation}
\label{overlapspincs}
\lg \widehat{\mathbf{r}}| \widehat{\mathbf{n}}\rg = \left( \frac{1 + \widehat{\mathbf{n}}\cdot  \widehat{\mathbf{r}}}{2}\right)^j\, \left( e^{\ii\,  A(\widehat{\mathbf{k}},\widehat{\mathbf{n}},\widehat{\mathbf{r}})}\right)^j\, , 
\end{equation}
The topological term $A(\widehat{\textbf{k}},\widehat{\mathbf{n}},\widehat{\mathbf{r}})$ appearing in the phase is the (symplectic) oriented area  of the geodesic spherical triangle with vertices $\widehat{\mathbf{k}},\widehat{\mathbf{n}},\widehat{\mathbf{r}}$.  In terms of the complex parameters  $\zeta=  \tan{\frac{\theta}{2}}\, e^{i\phi}$, $z = \tan{\frac{\eta}{2}}\, e^{i\psi}$ the conjugate of  this overlap reads as 
\begin{equation}
\label{overlapspincszeta}
\lg \zeta| z\rg= \overline{\lg z | \zeta\rg} = \frac{(1 + \bar{\zeta}z)^{2j}}{(1 + \vert z \vert^2)^j(1 + \vert \zeta \vert^2)^j}\,, 
\end{equation}
which is proportional to a holomorphic polynomial of degree $2j$ in the variable $z$. 
We observe from \eqref{overlapspincs} that the  orthogonality of two CS's holds for antipodal points: $\lg -\widehat{\mathbf{n}} | \widehat{\mathbf{n}}\rg = 0$, \textit{i.e.}, $z=-1/\bar{\zeta}$.
Therefore the SU$(2)$ spin coherent states  have just one zero on $\mathbb{S}^2$ or on the complex plane. Since they  are considered the closest to classical states in a certain sense, it has been conjectured that the more zeros the lower (or upper) symbol of a state has on $\mathbb{S}^2$ or on $\C$, the more quantum it is.
 
 The same conjecture can be extended to states on the complex plane ${z = b + \ii\omega}$ exhibiting Weyl-Heisenberg symmetry, as well as to states on the semi-discrete cylinder \cite{fabre23} with semi-discrete Weyl-Heisenberg symmetry. In this work, we outline promising directions for future research in this area, with potential applications in quantum optics and signal analysis. To make our discussion more concrete, we will focus on a specific example that is both manageable and numerically tractable.
 
\begin{itemize}
\item  Pick a finite set $\{z_i= b_i+ \ii\omega_i\}$ of $n$ points in the complex plane, \textit{e.g.}, located on the lattice $\Z^2$
\item Consider the polynomials $p_n(z) = \prod_{i=1}^n(z-z_i)$, a number $s\in (0,1)$ and the probability distribution
\begin{equation}
\label{distsn}
 \mathfrak{w}_{s,n}(\omega,b)= \frac{1}{\mathcal{N}}e^{\left[-(1-s)b^2-\left(\frac 1 s -1\right)\omega^2\right]}\, \vert p_n(b+\ii\omega)\vert^2\,
\end{equation}
with normalisation factor $\mathcal{N}$.
\item The exponential factor $e^{\left[-(1-s)b^2-\left(\frac 1 s -1\right)\omega^2\right]}$ is borrowed to the  example of holomorphic Hermite polynomials $H_n$ (see, \textit{e.g.}, \cite{gazsza11}) for which the orthogonality relations hold
\begin{equation*}
   \int_{\R^2}H_m(b+\ii \omega)\overline{H_n(b+\ii
\omega)}\exp\left[-(1-s)b^2-\left(\frac 1 s -1\right)\omega^2\right] 
\ud b\,  \ud \omega=b_n(s)\delta_{m,n}
   \end{equation*}
 where
   \begin{equation*}
   b_n(s)=\frac {\pi\sqrt s}{1-s}\left(2\frac{1+s}{1-s}\right)^n
n!\,.
   \end{equation*}
\end{itemize}

\begin{figure}[H]
\begin{center}
\includegraphics[width=5in]{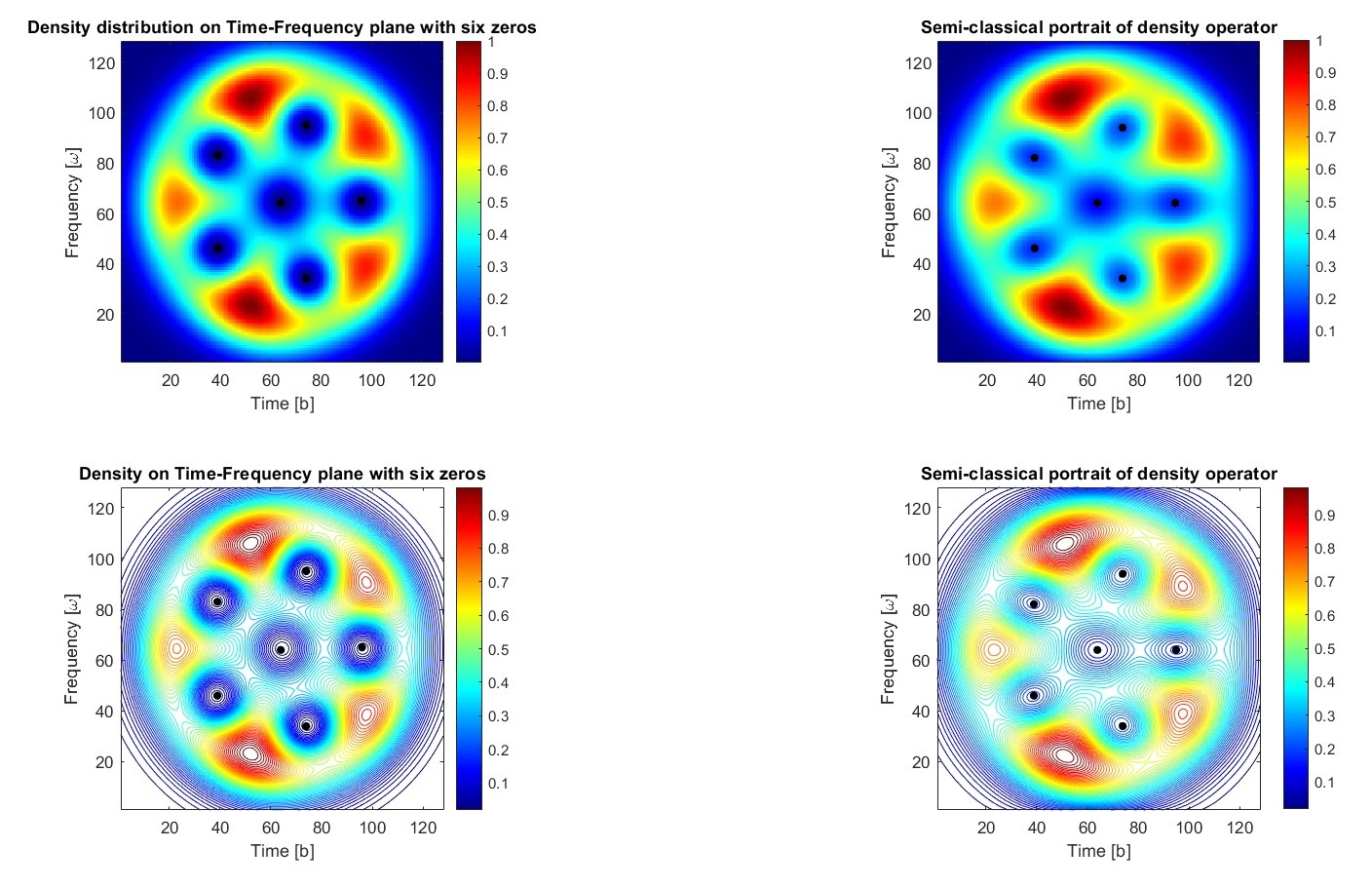}
\caption{Example of Majorana constellation in the time-frequency plane  $(\omega,b)$. In this figure we visualize both the density distribution $\mathfrak{w}_{s,n}(\omega,b)$ and its semi-classical portrait $\widecheck{\mathfrak{w}}^{\psi_{\mathrm{a}\mathrm{r}}}_{s,n}(\omega,b)$. In this particular visualization, $a=b=2$ and $s = 0.945$. The original  probability distribution is the expression \eqref{distsn} with polynomial $p_6(z)= z\prod_{m=1}^5\left(z-e^{\ii \frac{2\pi m}{5}}\right)$.  At the top are  shown the distributions $\mathfrak{w}_{s,n}$ (left) and $\widecheck{\mathfrak{w}}^{\psi_{\mathrm{a}\mathrm{r}}}_{s,n}$ (right) on the time-frequency plane. At the bottom are portrayed the contour lines, respectively. As $n = 6$, we expect to see 6 zeros.  Most interestingly, the conversion of $\mathfrak{w}_{s,n}(\omega,b)$ to the semi-classical portrait does not result in a major shift in these zeroes' positions. Some shift is expected, but overall, the original five-fold symmetry, the zeros,  and their positions are preserved under this transformation.} 
\label{pentamajo}
\end{center}
\end{figure}

\begin{itemize}
\item  Then study the functional properties of the states  
 \begin{equation}
\label{WHiqmapvp}
\rho^{\psi_{\mathrm{a}\mathrm{r}}}_{\mathfrak{w}_{s,n}}=\int_{\R^2}\,\rho_{\mathrm{a}}(\omega,b)\, \mathfrak{w}_{s,n}(\omega,b)\,\frac{\ud \omega\,\ud b}{2\pi}\,, \quad \rho_{\mathrm{a}}= |\psi_{\mathrm{a}}\rg\lg\psi_{\mathrm{a}}|\,.
\end{equation}
\item Next compare the two distributions on the time-frequency plane: 
\begin{equation}
\label{WHlocavervp}
\mathfrak{w}_{s,n}(\omega,b) \quad \mbox{versus}\quad  \widecheck{\mathfrak{w}}^{\psi_{\mathrm{a}\mathrm{r}}}_{s,n}(\omega,b)= \left(\mathfrak{w}_{s,n}\ast\mathsf{P}^{\mathrm{ar}}\right)(\omega,b)\,,
\end{equation}
\item particularly the locations of their respective  zeros.
\end{itemize}
In Fig. \ref{pentamajo} one can see an example with a polynomial having five zeros located at the vertices of a pentagon, plus one located  at the origin of the plane. It is interesting to observe the almost conservation of the original five-fold symmetry and of the locations of all zeroes. 

\section{Conclusion}
\label{conclu}
 In this contribution, a (partial) panorama of the Weyl-Heisenberg group(s) has been presented, ranging from the $3 \times 3$ matrix versions to the unitary irreducible representations and the resulting covariant integral representations.
 Our original approach was from the signal analysis perspective, with no Planck constant $\hbar$ involved.
 However, all the material presented is easily transposable to quantum mechanics, quantum optics, and quantum field theory.
  We have considered these questions in the context of their extensions to methods in physics involving other groups.
 The main points we would like to emphasize are the analysis and reconstruction of quantum states through classical and quantum probabilistic equations
 \begin{equation*}
\label{WHiqmapvp}
\rho^{\rho_{\mathrm{a}}}_{\mathfrak{w}}=\int_{\R^2}\,\rho_{\mathrm{a}}(\omega,b)\, \mathfrak{w}(\omega,b)\,\frac{\ud \omega\,\ud b}{2\pi}\,, \quad \rho_{\mathrm{a}}(\omega,b)= \mathcal{D}(\omega,b)\rho_a \mathcal{D}(-\omega,-b)\, .
\end{equation*}
 \begin{equation*}
\label{WHlocavervp}
\widecheck{\mathfrak{w}}^{\rho_{\mathrm{ar}}}(\omega,b)=\int_{\R^2}\mathfrak{w}(\omega^{\prime},b^{\prime})\,\mathrm{tr}(\rho_{\mathrm{r}}(\omega,b)\rho_{\mathrm{a}}(\omega^{\prime},b^{\prime}))\,\frac{\ud \omega^{\prime}\,\ud b^{\prime}}{2\pi}\,.
\end{equation*}
To finish this review, we mention another exploration to be made. It concerns the unitary dual (see \cite{kirillov62,kirillov04,lipsman74}) of the group SL$_{>}(n+1, \mathbb{R})$ of the unit upper real matrices and the  subdiagonal organisation  of their respective elements of the type $\left(z=\mathbf{x}^{(1)}, \mathbf{x}^{(2)}, \dotsc, \mathbf{x}^{(n-1)},\mathbf{x}^{(n)}\right)$, with $\mathbf{x}^{(k)} \in \R^k$, which generalises \eqref{GenWH3}. It deserves to implement the program of covariant (de-)quantisation integration which has described in the present paper in view of an application to signal analysis or quantum physics.

\section*{Acknowledgments}
J.-P. G, R. M, and A. Z thank Worcester Polytechnic Institute (WPI) for financial support. J.-P. G is grateful to the Physics Department of 
WPI for supporting his visit while finalizing this work.

\end{document}